\documentclass[]{article}
\usepackage[margin=1.5in]{geometry}
\usepackage[english]{babel}
\usepackage[utf8x]{inputenc}
\usepackage{amsmath}
\usepackage{graphicx}
\usepackage{setspace}
\usepackage{lineno}
\usepackage{cite}

\title{Experimental Observations of the Effects of Intermolecular Van der Waals Forces on Entropy}
\author{Matthew David Marko\\Marko Motors LLC\\Seaside Heights, NJ, USA\\mattdmarko@gmail.com}
\date{2 May 2022}

\begin{document}

\maketitle

\noindent Published in Nature Scientific Reports, {\bf{12}}, 7105, 2 May 2022.\\http://doi.org/10.1038/s41598-022-11093-z

\section*{Abstract}
\noindent An experimental effort was conducted to measure the change in internal energy of non-ideal carbon dioxide as its volume rapidly expanded with the sudden opening of a valve from one to two compressed gas cylinders.  This was achieved by measuring the mass heat capacity of the gas cylinders and the manifold-valve, and measuring the change in temperature from the sudden doubling of volume of the non-ideal carbon dioxide.  It was determined that an empirical equation for the change in internal energy of a non-ideal fluid was more accurate than previous methods used for estimating the change in internal energy by estimating the change in entropy.  With this empirical equation, a theoretical ideal Stirling cycle heat engine that exceeds the Carnot efficiency was realized by utilizing non-ideal carbon dioxide as a working fluid.  

\section{Introduction}

In an earlier publication \cite{MarkoAIP}, a theoretical Stirling cycle heat engine utilizing a real working fluid with significant intermolecular attractive (and repulsive) Van der Waals forces \cite{ClausiusOrig,1,2,3,4,StatThermo,Keesom2, KeesomOrig, LondonDispOrig, IntermolCermanic, RS_DispersionTemp, TempDepPhysRevB, Entropy01, Entropy02, Entropy03, Entropy04, Entropy05, Entropy06, Carnot_PRL,Carnot_PRX,Carnot_NJOP} was proposed.  The intermolecular attractive Van der Waals forces would both decrease the required work input during the cold isothermal compression, as well as reduce the work output recovered during the hot isothermal expansion.  If one were to look at empirical equations of state for real fluids, such as Redlich-Kwong \cite{RK1949} and Peng-Robinson \cite{PR1976}, it becomes clear that the intermolecular attractive force increases with decreasing temperatures.  Because of the temperature dependence of the Van der Waals forces, which increase in strength with decreasing temperature, the reduction in cold work input is greater than the loss of the hot work output; therefore, the ideal efficiency of this macroscopic heat engine could in theory exceed the Carnot efficiency ${\eta}_C$
\begin{eqnarray}
\label{eq:eqEffCarnot}
{\eta}_C&=&{1-{\frac{T_L}{T_H}}}.
\end{eqnarray}

Originally there were measurements of the enthalpy of vaporization of water \cite{NISTsteamHvEqu, NISTsteamDataTable}; when calculating for the change in internal energy during vaporization, which is isothermal expansion of a real fluid, it was observed that the change in specific internal energy ${{\delta}u}_{{\delta}T=0}$ (J/kg) during isothermal compression and expansion followed a distinct empirical equation
\begin{eqnarray}
\label{eq:eqDeltaU_isothermal_mytheory}
{{\delta}u}_{{\delta}T=0}&=&{{{a'}{\cdot}{{T}^{-0.25}}}{\cdot}{({\rho_1}-{\rho_2})}}, \\ 
{a'}&=&{\frac{0.21836}{9{\cdot}{({2^{\frac{1}{3}}-1})}}}}{\cdot}{\frac{{R_G^2}{\cdot}{T_C^{2.5}}}{{P_C}}.\nonumber 
\end{eqnarray}
where ${\rho}_1$ and ${\rho}_2$ (m$^3$/kg) represent the density, \emph{T} (K) represents the absolute temperature, $R_G$ (J/kg${\cdot}$K) represents the gas constant, $T_C$ (K) represents the critical temperature, and $P_C$ (Pa) represents the critical pressure.  This equation has been found to match well for numerous fluids \cite{MarkoAIP,New_NIST_R134a,New_NIST_N2,New_NIST_H2O,New_NIST_CH4, New_NIST_C2H6, New_NIST_C3H8, New_NIST_C4H10n, New_NIST_C4H10iso, NISTargon1, NISTargon2, ArgonCV, ArgonHvThesis, ArgonCriticalProp, nistXn, Beattie_1951_Xn, HvNobelGases, nistN2, nistAmmonia, NISTsteamHvEqu, NISTwaterVolDat1, NISTwaterCritProp, NISTwaterVolDat2, NISTsteamDataTable, GoffGratch1946,nistCxHy,BWR1940}.

For non-isothermal changes in specific internal energy of a fluid ${{\delta}u}$ (J/kg), one must include the specific intermolecular kinetic energy $u_{KE}={C_V}{\cdot}{R_G}{\cdot}{T}$ (J/kg) \cite{KinTheoryBornGreen1946}, 
\begin{eqnarray}
\label{eq:eqDeltaU_mytheory}
{{\delta}u}&=&{{C_V}{\cdot}{R_G}{\cdot}{{\delta}{T}}}-{{{a'}{\cdot}{{T}^{-0.25}}}{\cdot}{{\delta}{\rho}}}, \\ 
{a'}&=&{\frac{0.21836}{9{\cdot}{({2^{\frac{1}{3}}-1})}}}}{\cdot}{\frac{{R_G^2}{\cdot}{T_C^{2.5}}}{{P_C}}.\nonumber 
\end{eqnarray}
where \emph{T} (K) represents the absolute temperature, $R_G$ (J/kg${\cdot}$K) represents the gas constant, and ${C_V}$ is equal to the number of degrees of freedom of the molecule plus one half (ex. monatomic fluids $C_V=1.5$, diatomic fluids $C_V=2.5$, etc).  By integrating equation \ref{eq:eqDeltaU_mytheory} from a given density $\rho$ (kg/m$^3$) to infinitely low density (a true ideal gas) to find the intermolecular potential energy, and the temperature from absolute zero to the current temperature \emph{T}, one can calculate the total specific internal energy \emph{u} (J/kg) with equation \ref{eq:eqU_mytheory}
\begin{eqnarray}
\label{eq:eqU_mytheory}
{u}&=&{{C_V}{\cdot}{R_G}{\cdot}T}-{{{a'}{\cdot}{\rho}}{\cdot}{{T}^{-0.25}}},\\ \nonumber
{a'}&=&{\frac{0.21836}{9{\cdot}{({2^{\frac{1}{3}}-1})}}}}{\cdot}{\frac{{R_G^2}{\cdot}{T_C^{2.5}}}{{P_C}}.
\end{eqnarray}

Clausius' Theorem for the second law \cite{ClausiusOrig}
\begin{eqnarray}
\label{eq:eqClausius}
{{\oint}{\frac{{\delta}q}{T}}}&\leq&0,
\end{eqnarray}
states that any internally reversible thermodynamic cycle must generate a positive entropy ${\delta}{s_u}{\geq}0$ to the surrounding universe, where the change in specific entropy ${\delta}s$ (J/kg$\cdot$K) is defined as \cite{1,2,3,4,StatThermo}
\begin{eqnarray}
\label{eq:eqSideal}
{\delta}{s}&=&\frac{{\delta}q}{T},
\end{eqnarray}
where \emph{T} (K) is the absolute temperature, and ${\delta}q$ (J/kg) represent the heat transferred per unit mass.  If a fluid were to consistently follow equation \ref{eq:eqClausius}, then the change in specific internal energy ${{\delta}u}$ (J/kg) would consistently follow equation \ref{eq:eqdU_ideal} \cite{MarkoAIP,1,2,4}
\begin{eqnarray}
\label{eq:eqdU_ideal}
{\delta}{u}&=&{{C_V}{\cdot}{R_G}{{\delta}T}}+{\{{T{\cdot}{(\frac{{\partial}P}{{\partial}T}})}_V-{P}\}{\cdot}{{\delta}v}}.  
\end{eqnarray}
It should be noted that in most of the experimental measurements of the enthalpy of vaporization \cite{MarkoAIP,New_NIST_R134a,New_NIST_N2,New_NIST_H2O,New_NIST_CH4, New_NIST_C2H6, New_NIST_C3H8, New_NIST_C4H10n, New_NIST_C4H10iso, NISTargon1, NISTargon2, ArgonCV, ArgonHvThesis, ArgonCriticalProp, nistXn, Beattie_1951_Xn, HvNobelGases, nistN2, nistAmmonia, NISTsteamHvEqu, NISTwaterVolDat1, NISTwaterCritProp, NISTwaterVolDat2, NISTsteamDataTable, GoffGratch1946,nistCxHy,BWR1940} there is great similarity between equation \ref{eq:eqDeltaU_mytheory} and equation \ref{eq:eqdU_ideal}.  

Clausius' equation \ref{eq:eqClausius} makes intuitive sense for a reversible thermodynamic process utilizing an ideal-gas as its working fluid.  With a real-fluid, however, intermolecular Van-der-Waals forces impact the molecular behavior and thermodynamic properties.  In most published references and tables, the specific internal energy \emph{u} (J/kg) is often set to zero at an arbitrary point (often the triple-point), and calculated assuming equation \ref{eq:eqdU_ideal}, which was formulated for the purpose of holding equation \ref{eq:eqClausius} applicable.  

In contrast, equation \ref{eq:eqDeltaU_mytheory} is an empirical equation based on measurements of the change in internal energy for numerous different molecules during vaporization.  In addition, equation \ref{eq:eqDeltaU_mytheory} makes physical sense, as it includes both the internal kinetic energy of the molecules as defined with the Kinetic Gas Theory \cite{KinTheoryBornGreen1946}, as well as the intermolecular potential energy.  Lennard-Jones \cite{4,sph,spam}, a well-established approximation for the potential energy from intermolecular attractive and repulsive Van der Waals forces, assumes the attractive force is inverse proportional to the molecular distance to the sixth power $r^{-6}$ (m$^{-6}$); the attractive intermolecular force is thus inverse proportion to the specific volume squared $v^{-2}$ (m$^{-6}$), and this is observed in most empirical equations of state for a real fluid \cite{RK1949,PR1976}.  Integrating a potential force inverse-proportional to the specific volume squared would yield a potential energy inverse proportional to the specific volume \emph{v} (m$^3$/kg), or proportional to the density $\rho$ (kg/m$^3$), as observed in equations \ref{eq:eqDeltaU_mytheory} and \ref{eq:eqU_mytheory}.  
\begin{eqnarray}
{{\int}_{v}^{\infty}}{\frac{{\delta}v_0}{v_0^2}}={\frac{1}{v}}={\rho}.\nonumber
\end{eqnarray}

\section{Non-Ideal Stirling Cycle Heat Engine}

A Stirling cycle heat engine \cite{2,MarkoAIP} has isothermal (constant temperature) compression at a cold sink (Stage 1-2), isochoric (constant volume) heating to a hot temperature (Stage 2-3), isothermal expansion at the hot temperature source (Stage 3-4), and isochoric cooling back to the cold temperature (Stage 4-1).  For a true, ideal-gas Stirling engine to reach the Carnot efficiency (equation \ref{eq:eqEffCarnot}), it is necessary for all of the heat output from the isochoric cooling to go to the isochoric heating.  

As a demonstration, carbon dioxide will be the working fluid, with a cold temperature of 32$^{\circ}$C and a hot temperature of 82$^{\circ}$C.  The density will shift from 70 kg/m$^3$ to 700 kg/m$^3$.  Using equation \ref{eq:eqEffCarnot}, the Carnot efficiency ${\eta}_C$
\begin{eqnarray}
{\eta}_C={1-{\frac{32+273.15}{82+273.15}}}=14.08\%.\nonumber
\end{eqnarray}
The thermodynamic properties of this Stirling cycle engine are tabulated in Table \ref{tb:tbStirlingEng}, using both equation \ref{eq:eqU_mytheory} for the specific internal energy \emph{u} (J/kg) and the National Institute of Standards and Technology (NIST) Chemistry WebBook \cite{NIST_Webbook}.  The pressures used to calculate the specific work inputs and outputs $w={\int}P{\cdot}{\delta}v$ (J/kg) from NIST \cite{NIST_Webbook} are tabulated in Table \ref{tb:tbPnist}.  The efficiency is calculated as
\begin{eqnarray}
\label{eq:eqEffStirling}
{\eta}&=&-{\frac{{w_{12}}+{w_{34}}}{{q_{23}}+{q_{34}}+{q_{41}}}}.  
\end{eqnarray}
If the Stirling engine is utilizing an ideal-gas, then inherently ${q_{23}}=-{q_{41}}$; with a real fluid ${q_{23}}>-{q_{41}}$, and thus these values need to be included in equation \ref{eq:eqEffStirling}.  

The efficiency $\eta$ calculated with the values from NIST \cite{NIST_Webbook}, derived with equation \ref{eq:eqdU_ideal}, is 13.92\%, 
\begin{eqnarray}
{{\eta}_{NIST}}=-{\frac{{w_{12}}+{w_{34}}}{{q_{23}}+{q_{34}}+{q_{41}}}}=-{\frac{{69,149}-{105,515}}{{47,000}+{252,155}-{37,850}}}={\frac{36,367}{261,305}}=13.92\%.\nonumber
\end{eqnarray}
The efficiency calculated with equation \ref{eq:eqU_mytheory} is 16.89\%, 
\begin{eqnarray}
{{\eta}_{calc}}=-{\frac{{w_{12}}+{w_{34}}}{{q_{23}}+{q_{34}}+{q_{41}}}}=-{\frac{{69,149}-{105,515}}{{37,603}+{211,232}-{33,516}}}={\frac{36,367}{215,319}}=16.89\%,\nonumber
\end{eqnarray}
which actually exceeds the Carnot efficiency $\eta_C$ of 14.08\%!  

If one calculates the change in entropy ${\delta}s$ (J/kg${\cdot}{^{\circ}}$C) to the ambient universe with equation \ref{eq:eqSideal} throughout the full cycle
\begin{eqnarray}
{{\delta}s_{NIST}}={{\oint}{\frac{{\delta}q}{T}}}=({\frac{{q_{23}}+{q_{34}}+{q_{41}}}{T_H}}+{\frac{q_{12}}{T_L}})=({\frac{261,305}{82+273.15}}+{\frac{-224,939}{32+273.15}})=-1.38,\nonumber \\
{{\delta}s_{calc}}={{\oint}{\frac{{\delta}q}{T}}}=({\frac{{q_{23}}+{q_{34}}+{q_{41}}}{T_H}}+{\frac{q_{12}}{T_L}})=({\frac{215,319}{82+273.15}}+{\frac{-178,953}{32+273.15}})=19.83,\nonumber
\end{eqnarray}
it is noticed that the net total change in entropy per cycle with the NIST \cite{NIST_Webbook} internal energies ${{\delta}s_{NIST}}$ (J/kg${\cdot}{^{\circ}}$C), derived with equation \ref{eq:eqdU_ideal}, obeys Clausius' equation \ref{eq:eqClausius}; equation \ref{eq:eqdU_ideal} was in fact originally derived to ensure a thermodynamic cycle obeys Clausius' equation \ref{eq:eqClausius}.  If one assumes the internal energy of a real fluid can be determined with equation \ref{eq:eqU_mytheory}, an empirical equation based on previous measurements of the enthalpy of vaporization \cite{MarkoAIP,New_NIST_R134a, New_NIST_N2,New_NIST_H2O,New_NIST_CH4, New_NIST_C2H6, New_NIST_C3H8, New_NIST_C4H10n, New_NIST_C4H10iso, NISTargon1, NISTargon2, ArgonCV, ArgonHvThesis, ArgonCriticalProp, nistXn, Beattie_1951_Xn, HvNobelGases, nistN2, nistAmmonia, NISTsteamHvEqu, NISTwaterVolDat1, NISTwaterCritProp, NISTwaterVolDat2, NISTsteamDataTable, GoffGratch1946,nistCxHy,BWR1940}
then the net total change in entropy per cycle ${{\delta}s_{calc}}$ (J/kg${\cdot}{^{\circ}}$C) fails to obey Clausius' equation \ref{eq:eqClausius}.  

If the theoretical macroscopic Stirling cycle heat engine utilizing real fluids described is to exceed the Carnot efficiency, then equation \ref{eq:eqDeltaU_mytheory} must be the most accurate description of the change in internal energy for a non-ideal fluid.  If Clausius' (equation \ref{eq:eqClausius}) remains applicable in the presence of intermolecular Van der Waals forces, however, then equation \ref{eq:eqdU_ideal} would apply; inherently equation \ref{eq:eqdU_ideal} would have greater changes in internal potential energy during isothermal compression and expansion of a real fluid (observed in Table \ref{tb:tbStirlingEng}), and ensuring the ideal Stirling efficiency $\eta$ is less than or equal to the Carnot efficiency ${\eta}_C$ defined in equation \ref{eq:eqEffCarnot}.  It is thus desired to perform an experiment to determine which, equation \ref{eq:eqDeltaU_mytheory} or equation \ref{eq:eqdU_ideal}, is the most accurate definition of the change in specific internal energy ${\delta}{u}$ (J/kg) of a non-ideal fluid.

\section{Experimental Description}

\begin{figure}
\centering
\includegraphics[width=\linewidth]{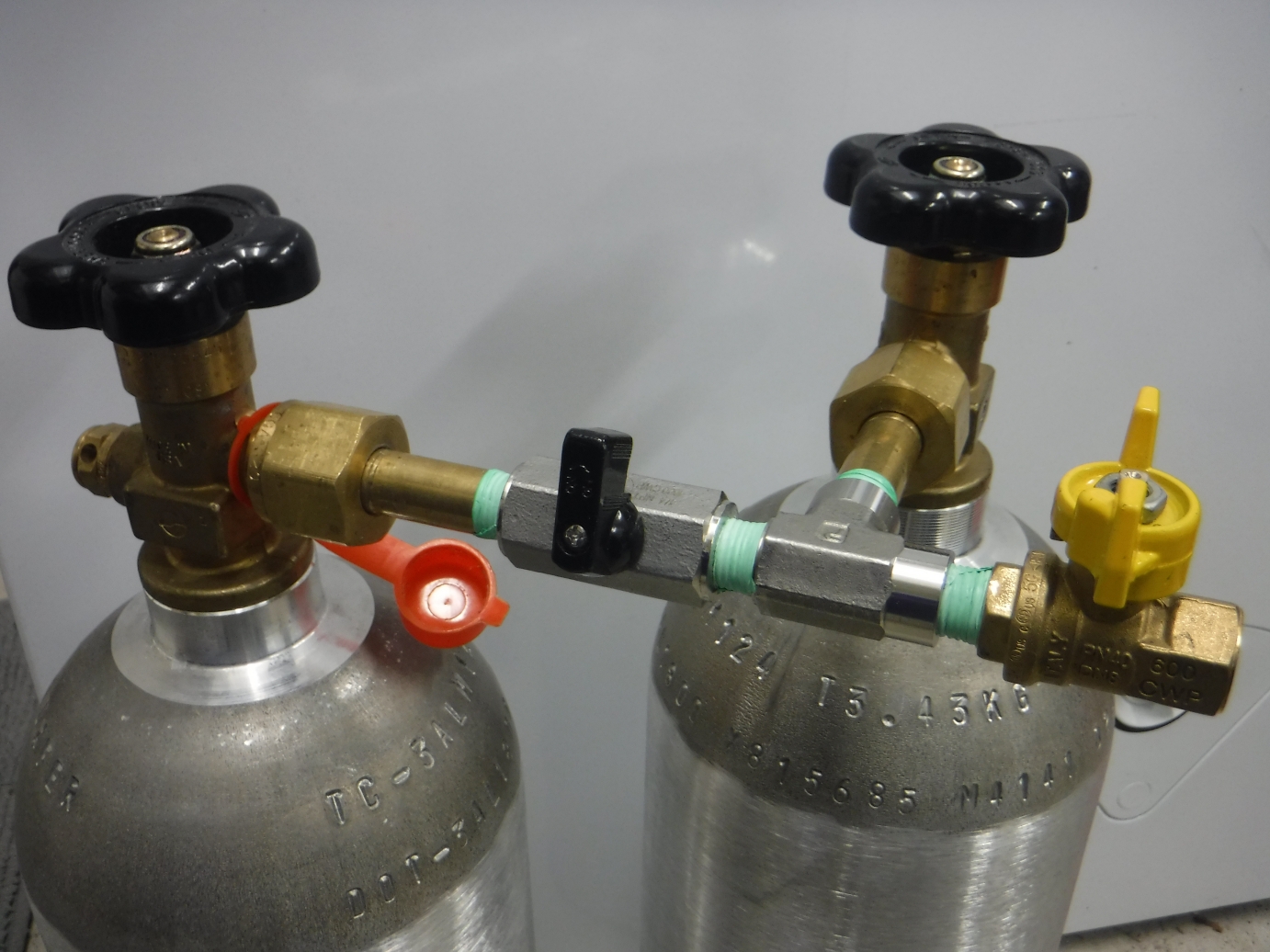}
\caption{The manifold to connect the two 3.4 liter CO$_2$ cylinders.  }
\label{fig:fig_Manifold}
\end{figure}

To determine if equation \ref{eq:eqDeltaU_mytheory} or equation \ref{eq:eqdU_ideal} is the most accurate definition of ${\delta}{u}$ (J/kg), a simple and easily repeatable experiment was performed.  Two compressed gas cylinders, manufactured by \emph{Luxfur}, with a volume of 3.4 liters each, and designed to hold 5 lbs of carbon dioxide (CO$_2$), were obtained.  They were connected by a manifold assembly (Figure \ref{fig:fig_Manifold}) which included adapters from the CGA-320 valve to NPT, a ball valve, a tee, and a separate ball valve (to bleed the small amount of CO$_2$ during disassembly).  Three calibrated thermocouples (Taylor \# 9940, Panel-Mount LCD Thermometer with Remote Probe; range  {-40}$^{\circ}$C to 150$^{\circ}$C) were used, with one attached via aluminum tape to each cylinder, and the third attached to the manifold.  

Before the experiment could commence, it was necessary to characterize the mass heat capacity $mC_P$ (J/$^{\circ}$C) of both the individual cylinders, as well as the manifold assembly.  First, the cylinder or manifold was left in a freezer (maintained at a temperature of -20$^{\circ}$C) for at least 24 hours, and the temperature inside the freezer was first measured and recorded as $T_{0,Sample}$ ($^{\circ}$C).  A mass (kg) of water was first weighed, and this water was poured into a bag resting inside an insulated chest.  A thermocouple was left in the bottom of the bag, and the temperature was then collected as the initial temperature of the water $T_{0,Water}$ ($^{\circ}$C).  The sample, either the cylinder or the manifold, was quickly moved into the water-filled bag, and insulating material was then piled on top of the bag to the limit that the insulated chest could be securely closed.  The water temperature, measured by the thermocouple, would quickly drop and later settle.  It was observed that the temperature would often settle after 15 minutes, but the final temperature $T_F$ ($^{\circ}$C) was collected after 60 minutes (minimal difference between the 15 minute measurements).  By using the NIST Chemistry WebBook \cite{NIST_Webbook} to determine the heat $Q_{water}$ (J) out of the mass of water from the change in temperature, the heat into the sample could be determined, and the mass heat capacity $mC_P$ (J/$^{\circ}$C)  could be estimated
\begin{eqnarray}
{mC_P}&=&{\frac{Q_{water}}{{T_F} - {T_{0,Sample}}}}.  
\end{eqnarray}
This was performed twice with both the cylinder and the manifold, and the results are tabulated in Table \ref{tb:tbExp_mCp}.  The averaged measured mass heat capacity of the cylinder is 2,524 J/${^{\circ}}$C, and the mass heat capacity of the manifold is 454 J/${^{\circ}}$C.  

\begin{figure}
\centering
\includegraphics[width=\linewidth]{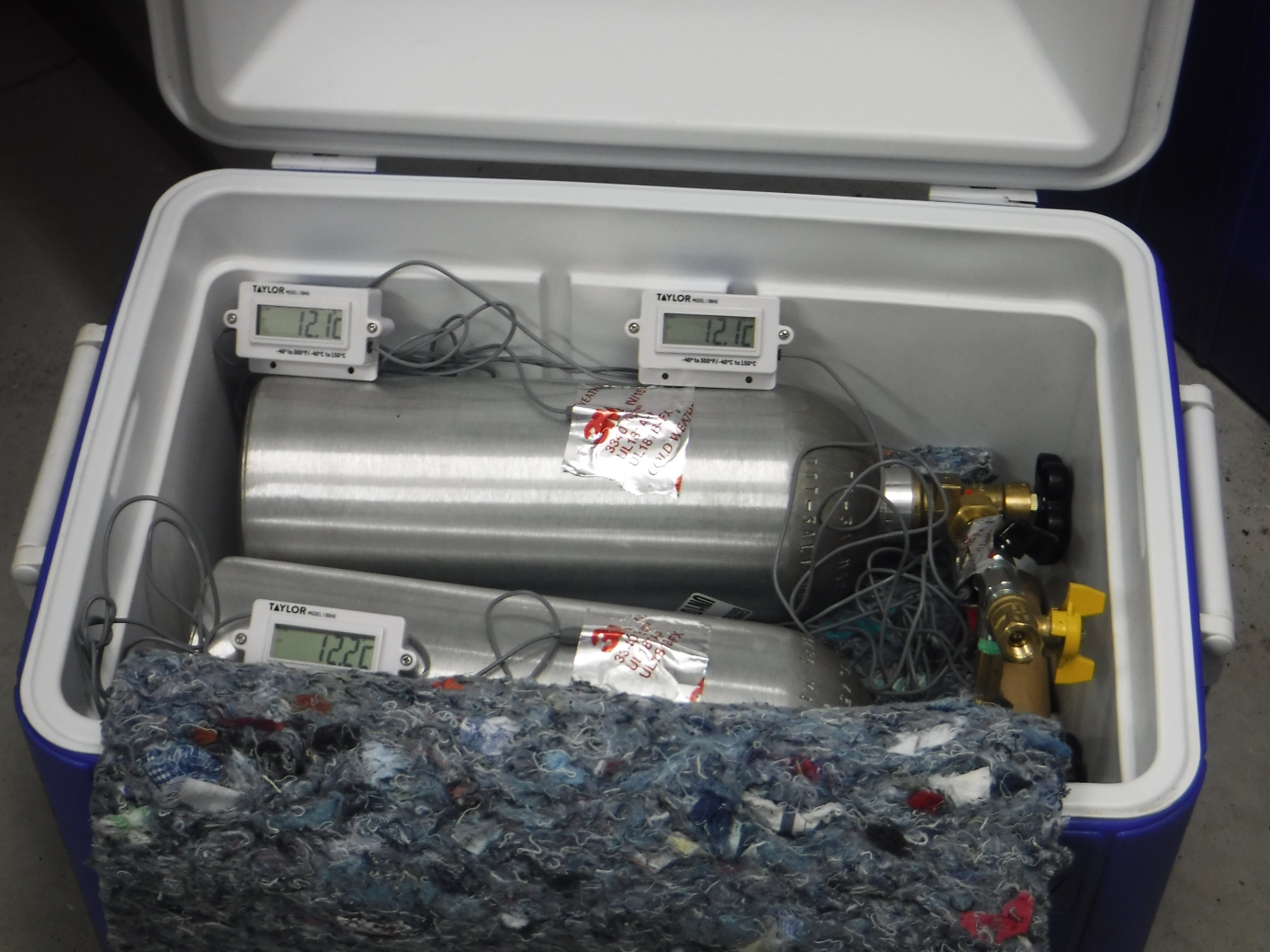}
\caption{The full experimental apparatus, with two CO$_2$ cylinders connected by the manifold (Figure \ref{fig:fig_Manifold}), inside the insulated chest, with the three thermocouples attached.  }
\label{fig:fig_Test_Setup}
\end{figure}

The process of the experiment was to have a mass of CO$_2$ in Cylinder 1, and leave Cylinder 2 empty.  The mass of CO$_2$ was determined by simply weighing Cylinder 1 before the experiment, and subtracting the measured mass of the empty cylinder (3,460.6 g).  The initial temperature on Cylinder 1, Cylinder 2, and the manifold (3) was recorded, and then the two-cylinder assembly was added to the insulated chest and thoroughly covered in insulation (Figure \ref{fig:fig_Test_Setup}).  To commence the experiment, the ball valve was suddenly opened, allowing for CO$_2$ to flow from Cylinder 1 to Cylinder 2, thus instantly doubling the volume and suddenly dropping the temperature due to the Joule–Thomson effect \cite{JouleThomson}.  Over time, the temperature of the cylinders and manifold would drop, as heat would flow from the aluminum cylinders and steel manifold into the cooler CO$_2$ until they reached thermal equilibrium, and the final temperature on Cylinder 1, Cylinder 2, and the manifold (3) was recorded.  The results of these temperature measurements are tabulated in Table \ref{tb:tbExp_Numbers1}.  

Afterwards, the remaining mass of CO$_2$ in Cylinder 1 and Cylinder 2 was recorded, and these results, along with the initial mass in Cylinder 1, are tabulated in Table \ref{tb:tbExp_Numbers2}.  In addition, with the known temperatures (Table \ref{tb:tbExp_Numbers1}) and densities (mass over the 3.4 liter volume), the specific internal energy \emph{u} (kJ/kg) values as determined by the NIST Chemistry WebBook \cite{NIST_Webbook} (which were derived by NIST with equation \ref{eq:eqdU_ideal}) were collected and tabulated in Table \ref{tb:tbExp_Numbers2}.  On average, less than 1\% of the CO$_2$ was lost in the disassembly of the manifold or due to leaking.  In addition, it was frequently observed that only a small portion of the mass (approximately 400g) would travel from Cylinder 1 to Cylinder 2; this is expected, as much of the liquid CO$_2$ (and thus most of the mass) would stay in the original Cylinder 1 rather than travel through the manifold.  It was also noticed that at times Cylinder 2 would experience a slight \emph{increase} in temperature; mainly due to the kinetic energy of expansion, and the low masses of CO$_2$ that made its way into Cylinder 2.

\section{Experimental Analysis}

Carbon dioxide (CO$_2$) \cite{CO2_NIST} has a critical temperature $T_C$ of 304.128 K; a critical pressure $P_C$ of 7,377,300 Pa; a critical specific density $\rho_C$ of 467.6 (kg/m$^3$), 3 degrees of freedom, a molar mass of 44 g/mole, and a Pitzer eccentric factor \cite{PitzerAcentric} of 0.228.  The specific gas constant $R_G$ for CO$_2$ is 188.924 J/kg${\cdot}{^{\circ}}$C; the ideal-gas specific heat at a constant volume $C_V=3.5{\cdot}{R_G}$; and the specific heat ratio ${\kappa}={\frac{C_P}{C_V}}={\frac{4.5}{3.5}}=1.28$.  For CO$_2$, the value of \emph{a'} as defined in equation \ref{eq:eqU_mytheory} is 728.46 {Pa}${\cdot}${K$^{0.25}$}$\cdot${{m$^6$}${\cdot}$kg$^{-2}$}.

The density of saturated liquid CO$_2$ $\rho_L$ (kg/m$^3$) and saturated gas CO$_2$ $\rho_G$ (kg/m$^3$) is defined with equation \ref{eq:eqRhoLG} \cite{CO2_NIST} 
\begin{eqnarray}
\label{eq:eqRhoLG}
{ln(\frac{\rho_L}{\rho_C})}&=&{{\Sigma}_{i=1}^4}{a_i}{\cdot}{(1-{\frac{T}{T_C}})^{t_i}}, \\ \nonumber
{ln(\frac{\rho_G}{\rho_C})}&=&{{\Sigma}_{i=1}^5}{b_i}{\cdot}{(1-{\frac{T}{T_C}})^{u_i}}, 
\end{eqnarray}
where $T_C$ is 304.128 K, $\rho_C$ is 467.6 (kg/m$^3$), and the values of $a_i$, $t_i$, $b_i$, and $u_i$ are tabulated in Table \ref{tb:tbRhoLG}. 

Table \ref{tb:tbExp_Dens} contains the tabulated densities $\rho$ (kg/m$^3$) of the CO$_2$, both before the experiment $\rho_0$ (kg/m$^3$), and after the experiment in Cylinder 1 $\rho_{F1}$ (kg/m$^3$) and Cylinder 2 $\rho_{F2}$ (kg/m$^3$).  In addition, the densities of a saturated liquid $\rho_L$ (kg/m$^3$) and a saturated gas $\rho_G$ (kg/m$^3$) for the experimentally measured CO$_2$ temperatures (Table \ref{tb:tbExp_Numbers1}) as determined with equation \ref{eq:eqRhoLG} is also tabulated in Table \ref{tb:tbExp_Dens}.  With the density $\rho$ (kg/m$^3$), saturated liquid density $\rho_L$ (kg/m$^3$), and saturated gas density $\rho_G$ (kg/m$^3$), the vapor quality \emph{X} was determined with equation \ref{eq:eqX}, and tabulated in Table \ref{tb:tbQual_Exp}.  
\begin{eqnarray}
\label{eq:eqX}
X&=&\frac{(1/{\rho})-(1/{\rho_L})}{(1/{\rho_G})-(1/{\rho_L})}=\frac{{v}-{v_L}}{{v_G}-{v_L}}.
\end{eqnarray}
Utilizing the density $\rho$ (kg/m$^3$), saturated liquid density $\rho_L$ (kg/m$^3$), and saturated gas density $\rho_G$ (kg/m$^3$) tabulated in Table \ref{tb:tbExp_Dens}, as well as the experimentally measured temperatures tabulated in Table \ref{tb:tbExp_Numbers1}, and the masses of CO$_2$ tabulated in Table \ref{tb:tbExp_Numbers2}, the internal energy \emph{U} (kJ) was calculated using the empirical equation \ref{eq:eqU_mytheory}, and tabulated in Table \ref{tb:tbU_CO2_xLG}.  

For qualities \emph{X} greater than 1, the CO$_2$ is treated as a vapor, and the internal energy is estimated solely with the empirical equation \ref{eq:eqU_mytheory}, and the final internal energy \emph{U} (kJ) was tabulated in Table \ref{tb:tbU_final_theory}.  For qualities \emph{X} less than 1 (there were no measurements at a greater density than the saturated liquid density), the internal energy of the liquid-vapor mixture \emph{U} (kJ) was calculated with equation \ref{eq:eqUfctX} from the quality \emph{X} (Table \ref{eq:eqX}), saturated liquid internal energy $U_L$ (kJ), and the saturated gas internal energy $U_G$ (kJ), both tabulated in Table \ref{tb:tbU_CO2_xLG}.  
\begin{eqnarray}
\label{eq:eqUfctX}
U&=&{{U_L}{\cdot}{(1-X)}}+{{U_G}{\cdot}{X}}.
\end{eqnarray}
All of these final internal energies \emph{U} (kJ) are tabulated in Table \ref{tb:tbU_final_theory}.  

\begin{figure}
\centering
\includegraphics[width=\linewidth]{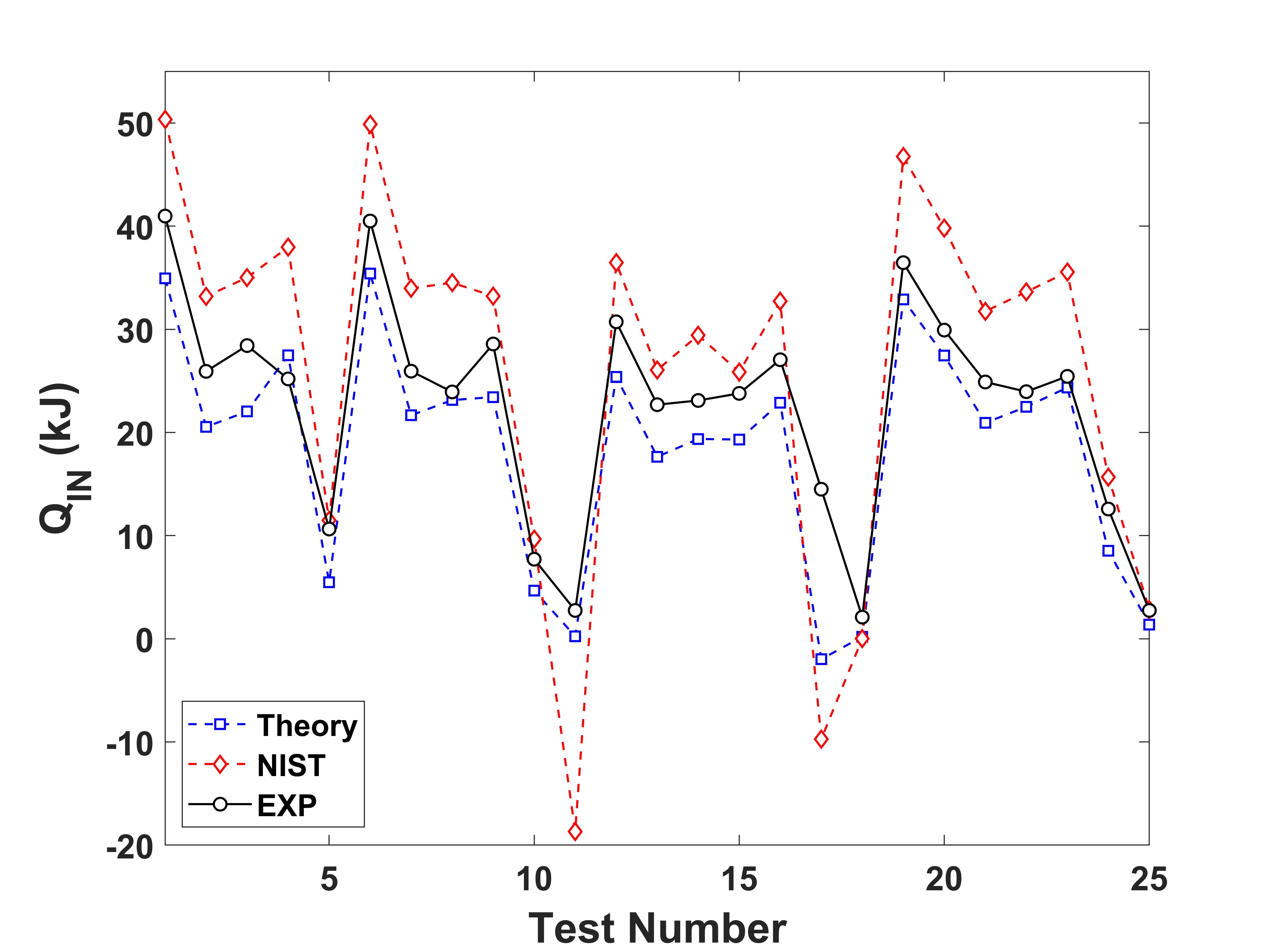}
\caption{The combined energy input \emph{Q} (kJ) into the CO$_2$, defined with equation \ref{eq:eqFindQ} and equation \ref{eq:eqFindQexp}, using the theory defined in equation \ref{eq:eqU_mytheory}, from NIST in Table \ref{tb:tbExp_Numbers2} \cite{NIST_Webbook}, and measured experimentally.  These plotted results are tabulated in Table \ref{tb:tbQin_combo}.}
\label{fig:fig_Q_in_figure}
\end{figure}

Finally, the estimated heat inputs were determined with equation \ref{eq:eqFindQ}, plotted in Figure \ref{fig:fig_Q_in_figure}, and tabulated in Table \ref{tb:tbQin_combo}.  These include the heat estimates $Q_{theory}$ (kJ) utilizing the internal energies derived from the empirical equation \ref{eq:eqU_mytheory} and tabulated in Table \ref{tb:tbU_final_theory}; as well as the heat estimates $Q_{NIST}$ (kJ) utilizing the specific internal energies collected from the NIST Chemistry WebBook \cite{NIST_Webbook} and tabulated in Table \ref{tb:tbExp_Numbers2}.  
\begin{eqnarray}
\label{eq:eqFindQ}
Q&=&{U_{F1}}+{U_{F2}}-{U_0}.
\end{eqnarray}
The values of $Q_{theory}$ (kJ) and $Q_{NIST}$ (kJ) are compared to the experimentally measured heat inputs $Q_{EXP}$ (kJ), determined by comparing the measured changes in temperature (Table \ref{tb:tbExp_Numbers1}) with the mass heat capacity tabulated in Table \ref{tb:tbExp_mCp}, as described in equation \ref{eq:eqFindQexp},
\begin{eqnarray}
\label{eq:eqFindQexp}
{Q_{EXP}}={{{mC_{P,Cylinder}}{\cdot}{({T_{1,0}}+{T_{2,0}}-{T_{1,F}}-{T_{2,F}})}}+{{mC_{P,Manifold}}{\cdot}{({T_{3,0}}-{T_{3,F}})}}}.
\end{eqnarray}

\section{Conclusion}

When analyzing the results of Table \ref{tb:tbQin_combo} and Figure \ref{fig:fig_Q_in_figure}, with 25 independent test results, the correlation between $Q_{Theory}$ (equation \ref{eq:eqDeltaU_mytheory}) and $Q_{EXP}$ is 0.9560; in excess of the correlation of 0.9229 between $Q_{NIST}$ (equation \ref{eq:eqdU_ideal}) and $Q_{EXP}$.  The average error between $Q_{Theory}$ and $Q_{EXP}$ is 29\%, less than the average error of 64\% between $Q_{NIST}$ and $Q_{EXP}$.  The median error between $Q_{Theory}$ and $Q_{EXP}$ is 17\%, less than the median error of 27\% between $Q_{NIST}$ and $Q_{EXP}$.  Finally, the standard deviation of the error between $Q_{Theory}$ and $Q_{EXP}$ is 29\%, less than the standard deviation of the error of 153\% between $Q_{NIST}$ and $Q_{EXP}$.  The experimental data suggests that equation \ref{eq:eqDeltaU_mytheory} is the most accurate definition of the change in internal energy of a real fluid ${\delta}{u}$ (J/kg), as compared to equation \ref{eq:eqdU_ideal}.  This effort provides an experimental justification to the possibility of the theoretical macroscopic Stirling cycle heat engine utilizing real fluids described earlier \cite{MarkoAIP} exceeding the Carnot efficiency.  

\section*{Data availability}
All data generated or analysed during this study are included in this published article [and its supplementary information files].

\clearpage

\begin{table}
\begin{center}
\begin{tabular}{ | c || c | c | c | c | c |}
  \hline
Stage & P (Pa) & T (K) & $\rho$ (kg/m$^3$) & $u_{NIST}$ (J/kg) & $u_{calc}$ (J/kg)\\
  \hline
  \hline
1 & 3,334,500 & 305.15 & 70 & 430,810 & 189,575\\
  \hline
2 & 8,650,400 & 305.15 & 700 & 275,020 & 79,771\\
  \hline
3 & 26,745,000 & 355.15 & 700 & 322,020 & 117,374\\
  \hline
4 & 4,132,200 & 355.15 & 70 & 468,660 & 223,091\\
  \hline
  \hline
Stages & w (J/kg) & ${\delta}u_{NIST}$ (J/kg) & $q_{NIST}$ (J/kg) & ${\delta}u_{calc}$ (J/kg) & $q_{calc}$ (J/kg)\\
  \hline
  \hline
12 & 69,149 & -155,790 & -224,939 & -109,804 & -178,953\\
  \hline
23 & 0 & 47,000 & 47,000 & 37,603 & 37,603\\
  \hline
34 & -105,515 & 146,640 & 252,155 & 105,717 & 211,232\\
  \hline
41 & 0 & -37,850 & -37,850 & -33,516 & -33,516\\
  \hline
\end{tabular}
\caption{The stages of the non-ideal Stirling cycle heat engine, as well as the specific work inputs and outputs \emph{w} (J/kg), specific heat inputs and outputs \emph{q} (J/kg), and specific internal energies \emph{u} (J/kg) from both \emph{NIST} \cite{NIST_Webbook} and calculated \emph{calc} with equation \ref{eq:eqU_mytheory}.}
\label{tb:tbStirlingEng}
\end{center}
\end{table}

\begin{table}
\begin{center}
\begin{tabular}{ | c || c | c || c | c |}
  \hline
$\rho$ (kg/m$^3$) & $P_{L}$ (MPa) & $P_{H}$ (MPa) & $u_{L}$ (kJ/kg) & $u_{H}$ (kJ/kg)\\
  \hline
  \hline
70 & 3.3345 & 4.1322 & 430.81 & 468.66\\
  \hline
88.146 & 3.5471 & 5.0367 & 428.9 & 463.45\\
  \hline
107.63 & 3.7598 & 5.9412 & 426.92 & 457.93\\
  \hline
128.64 & 3.9724 & 6.8457 & 424.86 & 452.08\\
  \hline
151.42 & 4.185 & 7.7502 & 422.73 & 445.85\\
  \hline
176.22 & 4.3977 & 8.6548 & 420.5 & 439.22\\
  \hline
203.31 & 4.6103 & 9.5593 & 418.18 & 432.14\\
  \hline
232.93 & 4.823 & 10.464 & 415.74 & 424.59\\
  \hline
265.24 & 5.0356 & 11.368 & 413.18 & 416.58\\
  \hline
300.17 & 5.2482 & 12.273 & 410.47 & 408.18\\
  \hline
337.28 & 5.4609 & 13.177 & 407.6 & 399.53\\
  \hline
375.61 & 5.6735 & 14.082 & 404.52 & 390.85\\
  \hline
413.83 & 5.8861 & 14.986 & 401.2 & 382.42\\
  \hline
450.56 & 6.0988 & 15.891 & 397.59 & 374.49\\
  \hline
484.73 & 6.3114 & 16.795 & 393.6 & 367.23\\
  \hline
515.82 & 6.524 & 17.7 & 389.12 & 360.69\\
  \hline
543.76 & 6.7367 & 18.604 & 383.94 & 354.85\\
  \hline
568.77 & 6.9493 & 19.509 & 377.72 & 349.63\\
  \hline
591.17 & 7.1619 & 20.413 & 369.68 & 344.96\\
  \hline
611.31 & 7.3746 & 21.318 & 357.28 & 340.75\\
  \hline
629.52 & 7.5872 & 22.222 & 303.52 & 336.94\\
  \hline
646.07 & 7.7999 & 23.127 & 289.34 & 333.46\\
  \hline
661.22 & 8.0125 & 24.031 & 283.9 & 330.26\\
  \hline
675.15 & 8.2251 & 24.936 & 280.23 & 327.31\\
  \hline
688.03 & 8.4378 & 25.84 & 277.38 & 324.57\\
  \hline
700 & 8.6504 & 26.745 & 275.02 & 322.02\\
  \hline
\end{tabular}
\caption{The pressures and specific internal energies versus the density $\rho$, taken from NIST \cite{NIST_Webbook}, and used to solve the work and heat inputs and outputs listed in Table \ref{tb:tbStirlingEng}.  The pressure $P_L$ (MPa) and specific internal energy $u_L$ (kJ/kg) is at 32$^{\circ}$C, and the pressure $P_H$ (MPa) and specific internal energy $u_H$ (kJ/kg) is at 82$^{\circ}$C.}
\label{tb:tbPnist}
\end{center}
\end{table}

\clearpage

\begin{table}
\begin{center}
\begin{tabular}{ | c || c | c | c | c | c |}
  \hline
Sample & Mass Water & $T_0$ \emph{Sample} & $T_0$ \emph{Water} & $T_F$ & $mC_P$\\
{} & (kg) & ($^{\circ}$C) & ($^{\circ}$C) & ($^{\circ}$C) & (J/$^{\circ}$C)\\
  \hline
  \hline
Cylinder & 4.01 & -16.1 & 23.7 & 18.4 & 2575.23\\
  \hline
Cylinder & 3.9 & -16.3 & 23.2 & 18 & 2473.09\\
  \hline
  \hline
Manifold & 1.6 & -20 & 24.7 & 21.8 & 464.56\\
  \hline
Manifold & 1.77 & -19.3 & 24.9 & 22.4 & 443.96\\
  \hline
\end{tabular}
\caption{The results of the effort to find the mass heat capacity $mC_P$ (J/$^{\circ}$C) of the cylinder and the manifold.  }
\label{tb:tbExp_mCp}
\end{center}
\end{table}

\begin{table}
\begin{center}
\begin{tabular}{ | c || c | c | c || c | c | c |}
  \hline
Test & $T_{1,0}$ & $T_{2,0}$ & $T_{3,0}$ & $T_{1,F}$ & $T_{2,F}$ & $T_{F,3}$ \\
Number & ($^{\circ}$C) &  ($^{\circ}$C) &  ($^{\circ}$C) &  ($^{\circ}$C) &  ($^{\circ}$C) & ($^{\circ}$C)\\
  \hline
  \hline
1-1 & 18.4 & 19.1 & 19.4 & 10.7 & 11.7 & 13.1\\
  \hline
1-2 & 18.2 & 18.7 & 18.3 & 7.9 & 19.9 & 11.8\\
  \hline
1-3 & 17.9 & 19.3 & 19.9 & 6.9 & 20.3 & 12.9\\
  \hline
1-4 & 18.4 & 19.2 & 18.7 & 8.4 & 20.3 & 12.7\\
  \hline
1-5 & 17.9 & 18.1 & 18.6 & 13.6 & 19.3 & 12.4\\
  \hline
  \hline
2-1 & 15.1 & 15.6 & 15.6 & 7.5 & 8.2 & 9.8\\
  \hline
2-2 & 14.7 & 15.2 & 16.0 & 4.8 & 16.1 & 8.9\\
  \hline
2-3 & 10.6 & 14.5 & 12.9 & 1.1 & 15.6 & 6.9\\
  \hline
2-4 & 14.5 & 15.5 & 16.2 & 3.4 & 16.7 & 8.3\\
  \hline
2-5 & 13.5 & 15.3 & 16.3 & 10.2 & 16.3 & 12.1\\
  \hline
2-6 & 13.1 & 13.3 & 13.2 & 11.4 & 14.2 & 11.6\\
  \hline
  \hline
3-1 & 10.9 & 10.0 & 10.3 & 3.1 & 6.6 & 4.9\\
  \hline
3-2 & 10.2 & 9.0 & 9.7 & 0.7 & 10.5 & 4.2\\
  \hline
3-3 & 7.6 & 8.9 & 10.2 & -1.7 & 10.2 & 3.8\\
  \hline
3-4 & 4.7 & 9.4 & 8.5 & -4.5 & 10.4 & 1.7\\
  \hline
3-5 & 11.2 & 11.5 & 11.4 & 0.4 & 12.7 & 5.2\\
  \hline
3-6 & 7.2 & 11.4 & 9.9 & 2.2 & 11.7 & 4.1\\
  \hline
3-7 & 8.2 & 11.2 & 11.0 & 7.0 & 12.0 & 8.6\\
  \hline
  \hline
4-1 & 10.9 & 11.8 & 12.2 & 4.5 & 4.8 & 6.4\\
  \hline
4-2 & 8.0 & 9.0 & 9.2 & 0.8 & 5.3 & 3.9\\
  \hline
4-3 & 9.8 & 11.1 & 11.0 & 0.5 & 11.6 & 5.1\\
  \hline
4-4 & 9.4 & 11.2 & 10.4 & -0.1 & 12.2 & 4.9\\
  \hline
4-5 & 8.0 & 11.2 & 10.9 & -1.6 & 12.0 & 3.8\\
  \hline
4-6 & 11.3 & 11.8 & 12.3 & 6.2 & 12.6 & 8.5\\
  \hline
4-7 & 11.6 & 12.1 & 12.2 & 9.9 & 13.0 & 10.6\\
  \hline
\end{tabular}
\caption{The measured temperature ($^{\circ}$C) of Cylinder \emph{1} and Cylinder \emph{2}, as well as the manifold temperature \emph{3}, both before \emph{0} and after \emph{F} the manifold valve was opened.}
\label{tb:tbExp_Numbers1}
\end{center}
\end{table}
 
\begin{table}
\begin{center}
\begin{tabular}{ | c || c | c | c || c | c | c |}
  \hline
{Test} & $mass_0$ &  $mass_{F1}$ & $mass_{F2}$ & $u_{0}$ & $u_{F1}$ & $u_{F1}$ \\
{Number} & (g) & (g) & (g) & (kJ/kg) & (kJ/kg) & (kJ/kg) \\
  \hline
  \hline
1-1 & 2026.7 & 1565.6 & 459.8 & 256.95 & 249.92 & 391.15\\
  \hline
1-2 & 1565.2 & 1200.2 & 364.1 & 272.42 & 258.99 & 408.49\\
  \hline
1-3 & 1199.1 & 848.4 & 350.8 & 292.83 & 285.54 & 410.19\\
  \hline
1-4 & 847.4 & 484.4 & 342.4 & 332.90 & 370.17 & 411.06\\
  \hline
1-5 & 483.8 & 246.9 & 236.3 & 394.34 & 415.90 & 421.39\\
  \hline
  \hline
2-1 & 1880.4 & 1385.4 & 491.0 & 251.91 & 247.99 & 366.59\\
  \hline
2-2 & 1385.4 & 1054.0 & 332.0 & 270.20 & 258.64 & 408.73\\
  \hline
2-3 & 1053.8 & 764.2 & 288.0 & 278.59 & 273.67 & 413.02\\
  \hline
2-4 & 764.0 & 444.7 & 305.1 & 329.22 & 357.58 & 412.07\\
  \hline
2-5 & 444.6 & 225.9 & 216.4 & 394.46 & 415.61 & 421.24\\
  \hline
2-6 & 226.3 & 110.9 & 109.5 & 417.76 & 258.16 & 431.34\\
  \hline
  \hline
3-1 & 1941.2 & 1582.5 & 349.8 & 238.88 & 227.90 & 398.84\\
  \hline
3-2 & 1582.2 & 1270.0 & 299.9 & 247.85 & 233.03 & 407.66\\
  \hline
3-3 & 1270.0 & 995.4 & 272.2 & 254.02 & 241.40 & 410.48\\
  \hline
3-4 & 995.8 & 738.6 & 238.5 & 262.85 & 255.59 & 414.36\\
  \hline
3-5 & 735.7 & 442.4 & 283.4 & 319.53 & 341.88 & 411.23\\
  \hline
3-6 & 442.6 & 197.7 & 182.7 & 380.36 & 412.78 & 421.50\\
  \hline
3-7 & 197.6 & 97.2 & 94.8 & 417.26 & 427.74 & 431.45\\
  \hline
  \hline
4-1 & 2233.5 & 1679.9 & 551.5 & 232.55 & 228.77 & 329.70\\
  \hline
4-2 & 1679.9 & 1356.8 & 321.0 & 238.32 & 229.54 & 401.02\\
  \hline
4-3 & 1356.8 & 1058.5 & 296.0 & 256.33 & 244.19 & 408.97\\
  \hline
4-4 & 1058.4 & 773.8 & 283.2 & 274.00 & 267.86 & 410.85\\
  \hline
4-5 & 773.5 & 503.8 & 268.1 & 299.86 & 311.51 & 412.35\\
  \hline
4-6 & 503.6 & 257.5 & 244.1 & 379.36 & 408.98 & 415.44\\
  \hline
4-7 & 257.2 & 128.7 & 126.6 & 413.23 & 426.21 & 428.62\\
  \hline
\end{tabular}
\caption{The measured mass (g) of CO$_2$ in the 3.4 liter cylinders, as well as the specific internal energy \emph{u} (kJ/kg) collected from the NIST webbook \cite{NIST_Webbook}.}
\label{tb:tbExp_Numbers2}
\end{center}
\end{table}

\begin{table}
\begin{center}
\begin{tabular}{ | c || c | c | c | c | c |}
  \hline
{ } & 1 & 2 & 3 & 4 & 5 \\
  \hline
$a_i$ & 1.9245108 & -0.62385555 & -0.32731127 & 0.39245142 & {-} \\
  \hline
$t_i$ & 0.340 & 0.5 & (10/6) & (11.6) & {-} \\
  \hline
  \hline
$b_i$ & -1.7074879 & -0.8227467 & -4.6008549 & -10.111178 & -29.742252\\
  \hline
$u_i$ & 0.340 & 0.5 & 1 & (7/3) & (14/3)\\
  \hline
\end{tabular}
\caption{Coefficient values for equation \ref{eq:eqRhoLG}.}
\label{tb:tbRhoLG}
\end{center}
\end{table}

\begin{table}
\begin{center}
\begin{tabular}{ | c || c | c | c || c | c | c || c | c | c |}
  \hline
{Test} & $\rho_{0}$ & $\rho_{0-L}$ & $\rho_{0-G}$ & $\rho_{F1}$ & $\rho_{F1-L}$ & $\rho_{F1-G}$ & $\rho_{F2}$ & $\rho_{F2-L}$ & $\rho_{F2-G}$ \\
{Num} & ($\frac{kg}{m^3}$) & ($\frac{kg}{m^3}$) & ($\frac{kg}{m^3}$) & ($\frac{kg}{m^3}$) & ($\frac{kg}{m^3}$) & ($\frac{kg}{m^3}$) & ($\frac{kg}{m^3}$) & ($\frac{kg}{m^3}$) & ($\frac{kg}{m^3}$)\\
  \hline
  \hline
1-1 & 596.09 & 789.83 & 182.35 & 460.47 & 855.86 & 138.39 & 135.22 & 848.19 & 143.19\\
  \hline
1-2 & 460.35 & 791.80 & 180.96 & 353.00 & 876.26 & 126.02 & 107.09 & 774.46 & 193.42\\
  \hline
1-3 & 352.68 & 794.72 & 178.89 & 249.53 & 883.22 & 121.94 & 103.18 & 770.16 & 196.58\\
  \hline
1-4 & 249.24 & 789.83 & 182.35 & 142.47 & 872.72 & 128.12 & 100.71 & 770.16 & 196.58\\
  \hline
1-5 & 142.29 & 794.72 & 178.89 & 72.62 & 832.99 & 152.94 & 69.50 & 780.74 & 188.86\\
  \hline
  \hline
2-1 & 553.06 & 820.32 & 161.31 & 407.47 & 879.06 & 124.37 & 144.41 & 874.14 & 127.28\\
  \hline
2-2 & 407.47 & 823.76 & 159.02 & 310.00 & 897.32 & 113.88 & 97.65 & 811.50 & 167.27\\
  \hline
2-3 & 309.94 & 856.61 & 137.92 & 224.76 & 920.75 & 101.12 & 84.71 & 815.95 & 164.25\\
  \hline
2-4 & 224.71 & 825.47 & 157.89 & 130.79 & 906.39 & 108.85 & 89.74 & 806.04 & 171.01\\
  \hline
2-5 & 130.76 & 833.81 & 152.40 & 66.44 & 859.61 & 136.07 & 63.65 & 809.69 & 168.51\\
  \hline
2-6 & 66.56 & 837.08 & 150.29 & 32.62 & 850.51 & 141.72 & 32.21 & 828.00 & 156.21\\
  \hline
  \hline
3-1 & 570.94 & 854.34 & 139.33 & 465.44 & 908.30 & 107.80 & 102.88 & 885.27 & 120.75\\
  \hline
3-2 & 465.35 & 859.61 & 136.07 & 373.53 & 923.18 & 99.84 & 88.21 & 857.36 & 137.46\\
  \hline
3-3 & 373.53 & 878.37 & 124.78 & 292.76 & 937.45 & 92.52 & 80.06 & 859.61 & 136.07\\
  \hline
3-4 & 292.88 & 897.98 & 113.51 & 217.24 & 953.40 & 84.68 & 70.15 & 858.11 & 136.99\\
  \hline
3-5 & 216.38 & 852.05 & 140.76 & 130.12 & 925.00 & 98.89 & 83.35 & 840.30 & 148.21\\
  \hline
3-6 & 130.18 & 881.15 & 123.15 & 58.15 & 913.96 & 104.74 & 53.74 & 848.19 & 143.19\\
  \hline
3-7 & 58.12 & 874.14 & 127.28 & 28.59 & 882.53 & 122.34 & 27.88 & 845.85 & 144.67\\
  \hline
  \hline
4-1 & 656.91 & 854.34 & 139.33 & 494.09 & 899.29 & 112.78 & 162.21 & 897.32 & 113.88\\
  \hline
4-2 & 494.09 & 875.56 & 126.44 & 399.06 & 922.58 & 100.16 & 94.41 & 894.02 & 115.74\\
  \hline
4-3 & 399.06 & 862.58 & 134.25 & 311.32 & 924.40 & 99.21 & 87.06 & 848.97 & 142.70\\
  \hline
4-4 & 311.29 & 865.51 & 132.46 & 227.59 & 928.01 & 97.33 & 83.29 & 844.28 & 145.67\\
  \hline
4-5 & 227.50 & 875.56 & 126.44 & 148.18 & 936.87 & 92.81 & 78.85 & 845.85 & 144.67\\
  \hline
4-6 & 148.12 & 851.28 & 141.24 & 75.74 & 887.99 & 119.18 & 71.79 & 841.10 & 147.70\\
  \hline
4-7 & 75.65 & 848.97 & 142.70 & 37.85 & 861.84 & 134.70 & 37.24 & 837.89 & 149.76\\
  \hline
\end{tabular}
\caption{The calculated density of CO$_2$ in the 3.4 liter cylinders, taken from the mass tabulated in Table \ref{tb:tbExp_Numbers2}.  The densities of a saturated liquid and a saturated gas are defined with equation \ref{eq:eqRhoLG}.}
\label{tb:tbExp_Dens}
\end{center}
\end{table}

\begin{table}
\begin{center}
\begin{tabular}{ | c || c | c | c |}
  \hline
{Test} & $X_{0}$ & $X_{F1}$ & $X_{F2}$\\
  \hline
  \hline
1-1 & 0.0976 & 0.1656 & 1.0709\\
  \hline
1-2 & 0.2133 & 0.2490 & 2.0746\\
  \hline
1-3 & 0.3641 & 0.4068 & 2.2155\\
  \hline
1-4 & 0.6511 & 0.8820 & 2.2782\\
  \hline
1-5 & 1.3319 & 2.3548 & 3.2654\\
  \hline
  \hline
2-1 & 0.1183 & 0.1907 & 0.8611\\
  \hline
2-2 & 0.2444 & 0.2754 & 1.8982\\
  \hline
2-3 & 0.3385 & 0.3820 & 2.1758\\
  \hline
2-4 & 0.6323 & 0.8093 & 2.1497\\
  \hline
2-5 & 1.2025 & 2.2451 & 3.0805\\
  \hline
2-6 & 2.5332 & 5.0139 & 5.7458\\
  \hline
  \hline
3-1 & 0.0967 & 0.1281 & 1.2011\\
  \hline
3-2 & 0.1593 & 0.1784 & 1.6650\\
  \hline
3-3 & 0.2238 & 0.2411 & 1.8312\\
  \hline
3-4 & 0.2989 & 0.3303 & 2.1340\\
  \hline
3-5 & 0.5814 & 0.7313 & 1.9447\\
  \hline
3-6 & 0.9372 & 1.9050 & 3.0028\\
  \hline
3-7 & 2.3928 & 4.8072 & 6.0527\\
  \hline
  \hline
4-1 & 0.0586 & 0.1176 & 0.6588\\
  \hline
4-2 & 0.1303 & 0.1598 & 1.2595\\
  \hline
4-3 & 0.2141 & 0.2368 & 1.7682\\
  \hline
4-4 & 0.3217 & 0.3606 & 1.9050\\
  \hline
4-5 & 0.4808 & 0.5853 & 2.0069\\
  \hline
4-6 & 0.9443 & 1.6626 & 2.2824\\
  \hline
4-7 & 2.0654 & 4.0326 & 4.6798\\
  \hline
\end{tabular}
\caption{The calculated vapor quality solved with equation \ref{eq:eqX}, utilizing the density $\rho$ (kg/m$^3$), saturated liquid density $\rho_L$ (kg/m$^3$), and saturated gas density $\rho_G$ (kg/m$^3$) tabulated in Table \ref{tb:tbExp_Dens}.}
\label{tb:tbQual_Exp}
\end{center}
\end{table}

\begin{table}
\begin{center}
\begin{tabular}{ | c || c | c | c || c | c | c || c | c | c |}
  \hline
{Test} & $U_{0x}$ & $U_{0-L}$ & $U_{0-G}$ & $U_{F1-x}$ & $U_{F1-L}$ & $U_{F1-G}$ & $U_{F2-x}$ & $U_{F2-L}$ & $U_{F2-G}$ \\
{Num} & (kJ) & (kJ) & (kJ) & (kJ) & (kJ) & (kJ) & (kJ) & (kJ) & (kJ)\\
  \hline
  \hline
1-1 & 177.74 & 108.52 & 325.56 & 165.91 & 56.05 & 255.40 & 75.57 & 17.45 & 74.92\\
  \hline
1-2 & 174.49 & 83.02 & 251.60 & 147.67 & 35.93 & 196.14 & 63.69 & 20.91 & 58.15\\
  \hline
1-3 & 156.19 & 62.70 & 192.94 & 119.41 & 23.67 & 138.68 & 61.70 & 20.52 & 55.93\\
  \hline
1-4 & 126.13 & 45.37 & 136.12 & 77.91 & 15.00 & 79.14 & 60.37 & 20.03 & 54.59\\
  \hline
1-5 & 80.97 & 25.30 & 77.84 & 43.64 & 10.41 & 40.13 & 42.80 & 13.20 & 37.83\\
  \hline
  \hline
2-1 & 174.55 & 85.70 & 304.78 & 156.63 & 40.34 & 226.43 & 78.73 & 15.00 & 80.23\\
  \hline
2-2 & 163.86 & 61.86 & 224.73 & 135.42 & 24.98 & 172.30 & 57.77 & 15.91 & 53.69\\
  \hline
2-3 & 139.75 & 37.50 & 171.92 & 107.84 & 12.63 & 124.75 & 50.68 & 13.46 & 46.63\\
  \hline
2-4 & 114.95 & 33.76 & 123.98 & 70.93 & 9.32 & 72.67 & 53.64 & 15.06 & 49.26\\
  \hline
2-5 & 73.98 & 18.64 & 72.27 & 39.66 & 7.85 & 36.87 & 38.99 & 10.47 & 34.98\\
  \hline
2-6 & 40.17 & 9.29 & 36.81 & 20.22 & 4.14 & 18.08 & 20.18 & 4.76 & 17.78\\
  \hline
  \hline
3-1 & 167.94 & 70.32 & 316.61 & 157.46 & 32.23 & 258.59 & 58.30 & 9.55 & 57.18\\
  \hline
3-2 & 165.71 & 54.96 & 258.22 & 145.02 & 20.02 & 207.26 & 51.55 & 10.61 & 48.93\\
  \hline
3-3 & 151.34 & 37.24 & 207.56 & 126.37 & 11.20 & 162.14 & 47.13 & 9.45 & 44.42\\
  \hline
3-4 & 130.91 & 23.40 & 162.78 & 102.34 & 4.50 & 119.95 & 41.75 & 8.39 & 38.92\\
  \hline
3-5 & 110.09 & 27.13 & 119.96 & 69.71 & 6.72 & 72.18 & 49.38 & 11.38 & 46.13\\
  \hline
3-6 & 71.79 & 12.62 & 72.34 & 33.94 & 3.68 & 32.29 & 32.67 & 6.93 & 29.77\\
  \hline
3-7 & 34.72 & 6.04 & 32.29 & 17.51 & 2.73 & 15.89 & 17.41 & 3.66 & 15.44\\
  \hline
  \hline
4-1 & 159.16 & 80.91 & 364.28 & 160.29 & 38.82 & 274.61 & 85.40 & 13.07 & 90.16\\
  \hline
4-2 & 164.64 & 50.64 & 274.52 & 148.83 & 21.64 & 221.44 & 53.70 & 7.93 & 52.48\\
  \hline
4-3 & 157.68 & 45.98 & 221.50 & 132.51 & 16.28 & 172.72 & 51.16 & 11.17 & 48.24\\
  \hline
4-4 & 139.20 & 34.98 & 172.83 & 108.15 & 11.02 & 126.21 & 49.25 & 11.06 & 46.12\\
  \hline
4-5 & 112.49 & 23.32 & 126.40 & 77.07 & 5.76 & 82.07 & 46.80 & 10.35 & 43.67\\
  \hline
4-6 & 81.49 & 18.68 & 82.10 & 44.09 & 6.82 & 42.10 & 43.02 & 9.75 & 39.73\\
  \hline
4-7 & 44.98 & 9.71 & 41.92 & 23.22 & 4.39 & 21.01 & 23.12 & 5.17 & 20.60\\
  \hline
\end{tabular}
\caption{The calculated internal energy \emph{U} (kJ), solved with equation \ref{eq:eqU_mytheory}, utilizing the measured temperature
in Table \ref{tb:tbExp_Numbers1}, and the densities tabulated in Table \ref{tb:tbExp_Dens}.}
\label{tb:tbU_CO2_xLG}
\end{center}
\end{table}

\begin{table}
\begin{center}
\begin{tabular}{ | c || c | c | c |}
  \hline
{Test} & $U_{0}$ & $U_{F1}$ & $U_{F2}$\\
{Num} & (kJ) & (kJ) & (kJ)\\
  \hline
  \hline
1-1 & 129.69 & 89.06 & 75.57\\
  \hline
1-2 & 118.97 & 75.82 & 63.69\\
  \hline
1-3 & 110.12 & 70.46 & 61.70\\
  \hline
1-4 & 104.46 & 71.57 & 60.37\\
  \hline
1-5 & 80.97 & 43.64 & 42.80\\
  \hline
  \hline
2-1 & 111.61 & 75.84 & 71.17\\
  \hline
2-2 & 101.66 & 65.55 & 57.77\\
  \hline
2-3 & 83.00 & 55.46 & 50.68\\
  \hline
2-4 & 90.81 & 60.59 & 53.64\\
  \hline
2-5 & 73.98 & 39.66 & 38.99\\
  \hline
2-6 & 40.17 & 20.22 & 20.18\\
  \hline
  \hline
3-1 & 94.15 & 61.24 & 58.30\\
  \hline
3-2 & 87.34 & 53.43 & 51.55\\
  \hline
3-3 & 75.36 & 47.59 & 47.13\\
  \hline
3-4 & 65.07 & 42.64 & 41.75\\
  \hline
3-5 & 81.09 & 54.60 & 49.38\\
  \hline
3-6 & 68.60 & 33.94 & 32.67\\
  \hline
3-7 & 34.72 & 17.51 & 17.41\\
  \hline
  \hline
4-1 & 97.51 & 66.55 & 63.85\\
  \hline
4-2 & 79.81 & 53.57 & 53.70\\
  \hline
4-3 & 83.56 & 53.32 & 51.16\\
  \hline
4-4 & 79.33 & 52.56 & 49.25\\
  \hline
4-5 & 72.88 & 50.42 & 46.80\\
  \hline
4-6 & 78.57 & 44.09 & 43.02\\
  \hline
4-7 & 44.98 & 23.22 & 23.12\\
  \hline
\end{tabular}
\caption{The calculated internal energies \emph{U} (kJ), solved with equation \ref{eq:eqU_mytheory}, adjusting for mixed liquid-vapor, using the internals energies \emph{U} (kJ) tabulated in Table \ref{tb:tbU_CO2_xLG}, with the qualities \emph{X} tabulated in Table \ref{tb:tbQual_Exp}, solved with equation \ref{eq:eqUfctX}.}
\label{tb:tbU_final_theory}
\end{center}
\end{table}

\begin{table}
\begin{center}
\begin{tabular}{ | c || c | c | c |}
  \hline
{Test} & $Q_{Theory}$ & $Q_{NIST}$ & $Q_{EXP}$\\
Num & (kJ) & (kJ) & (kJ)\\
  \hline
  \hline
1-1 & 34.9442 (14.71\%) & 50.3454 (22.88\%) & 40.9726\\
  \hline
1-2 & 20.5379 (20.76\%) & 33.1792 (28.01\%) & 25.9194\\
  \hline
1-3 & 22.0339 (22.47\%) & 35.0143 (23.21\%) & 28.4180\\
  \hline
1-4 & 27.4844 (9.12\%) & 37.9578 (50.70\%) & 25.1876\\
  \hline
1-5 & 5.4758 (48.53\%) & 11.4785 (7.89\%) & 10.6392\\
  \hline
  \hline
2-1 & 35.3962 (12.59\%) & 49.8695 (23.16\%) & 40.4932\\
  \hline
2-2 & 21.6599 (16.50\%) & 33.9698 (30.96\%) & 25.9394\\
  \hline
2-3 & 23.1390 (3.29\%) & 34.5102 (44.24\%) & 23.9256\\
  \hline
2-4 & 23.4268 (18.01\%) & 33.2143 (16.24\%) & 28.5742\\
  \hline
2-5 & 4.6671 (39.48\%) & 9.6657 (25.33\%) & 7.7120\\
  \hline
2-6 & 0.2402 (91.25\%) & -18.6774 (780.27\%) & 2.7456\\
  \hline
  \hline
3-1 & 25.3884 (17.36\%) & 36.4521 (18.66\%) & 30.7204\\
  \hline
3-2 & 17.6409 (22.25\%) & 26.0571 (14.84\%) & 22.6890\\
  \hline
3-3 & 19.3648 (16.16\%) & 29.4168 (27.36\%) & 23.0976\\
  \hline
3-4 & 19.3141 (18.79\%) & 25.8576 (8.72\%) & 23.7840\\
  \hline
3-5 & 22.8826 (15.39\%) & 32.7121 (20.95\%) & 27.0452\\
  \hline
3-6 & -1.9849 (113.69\%) & -9.7327 (167.14\%) & 14.4960\\
  \hline
3-7 & 0.1985 (90.54\%) & 0.0272 (98.70\%) & 2.0992\\
  \hline
  \hline
4-1 & 32.8876 (9.79\%) & 46.7398 (28.21\%) & 36.4548\\
  \hline
4-2 & 27.4498 (8.25\%) & 39.8135 (33.08\%) & 29.9178\\
  \hline
4-3 & 20.9232 (15.94\%) & 31.7417 (27.53\%) & 24.8898\\
  \hline
4-4 & 22.4886 (6.11\%) & 33.6212 (40.37\%) & 23.9510\\
  \hline
4-5 & 24.3473 (4.28\%) & 35.5481 (39.76\%) & 25.4346\\
  \hline
4-6 & 8.5325 (32.17\%) & 15.6756 (24.62\%) & 12.5784\\
  \hline
4-7 & 1.3649 (50.29\%) & 2.8338 (3.21\%) & 2.7456\\
  \hline
\end{tabular}
\caption{The combined energy input \emph{Q} (kJ) into the CO$_2$, defined with equation \ref{eq:eqFindQ} and equation \ref{eq:eqFindQexp}, using the theory defined in equation \ref{eq:eqU_mytheory}, from NIST in Table \ref{tb:tbExp_Numbers2} \cite{NIST_Webbook}, and measured experimentally.  The parentheses represent the percent (\%) error with $Q_{EXP}$.  These tabulated results are plotted in Figure \ref{fig:fig_Q_in_figure}.}
\label{tb:tbQin_combo}
\end{center}
\end{table}

\clearpage

\clearpage



\end{document}